\documentclass[AMA,STIX1COL,sort&compress,square]{WileyNJD-v2}

\usepackage{moreverb}
\newcommand\BibTeX{{\rmfamily B\kern-.05em \textsc{i\kern-.025em b}\kern-.08em
T\kern-.1667em\lower.7ex\hbox{E}\kern-.125emX}}
\articletype{Research Article}

\received{---}
\revised{---}
\accepted{---}
\DeclareUnicodeCharacter{2212}{-}
\usepackage{hyperref}
\hypersetup{
    colorlinks,
    citecolor=blue,
    filecolor=blue,
    linkcolor=blue,
    urlcolor=blue}    
\begin{document}

\title{Self-organization of earth's inner magnetospheric multi-ion plasma}
\author{Usman Shazad*}
\author{M. Iqbal}
\address{\orgdiv{Department of Physics}, \orgname{University of Engineering and Technology Lahore 54890}, \orgaddress{\country{Pakistan.}}}
\corres{*Usman Shazad, Department of Physics, University of Engineering and Technology Lahore 54890, Pakistan.\\ 
\email{usmangondle@gmail.com}\\https://orcid.org/0000-0001-7151-8542}
\abstract[Abstract]{The self-organization of a magnetized multi-ion plasma, composed of inertialess electrons and inertial H$^{+}$, He$^{+}$, and O$^{+}$ ions, leads to the formation of quadruple Beltrami (QB) field structures. The QB self-organized state is a linear combination of four single Beltrami fields, and it is a non-force-free state that shows strong magnetofluid coupling. Moreover, the QB state is characterized by four relaxed state structures of different length scales. The investigation reveals that the generalized helicities of plasma species and the densities of ion species have a significant impact on the characteristics of the self-organized vortices in the QB state. The study also highlights the potential consequences of QB field structures on earth's inner magnetosphere, including diamagnetic and paramagnetic trends as well as heating effects resulting from disparate length scales.}

\maketitle
\section{Introduction}\label{S1}
Multi-ion plasmas are prevalent in space and astrophysical objects. The ionosphere and magnetosphere of the earth, the solar wind, the bow shock in front of the magnetopause boundary layers, the heliosphere, the magnetosphere of Saturn, and cometary tails all contain multi-ion plasmas \cite{Stasiewicz2004,Melin2009,Hultqvist2012,Schwenn2012,Borovsky2018}.  In recent years, extensive research has been conducted on multi-ion plasmas in order to investigate various nonlinear structures. For instance, several researchers have studied ion cyclotron waves \cite{Gomberoff1983,Horne1993,Gomberoff1995,Hu2010,Patel2011}, kinetic Alfv\'{e}n waves \cite{Moya2015,Tamrakar2018,Tamrakar2018a,Tamrakar2019a,Moya2022}, electrostatic ion-cyclotron waves \cite{Raikwar2017}, inertial Alfv\'{e}n waves \cite{Tamrakar2019}, and oscillitons \cite{Sauer2022} in multi-ion plasmas containing three positive ion species that occur in various regions of the earth's magnetosphere.

The objective of this study is to investigate the magnetic self-organized state of a multi-ion plasma, consisting of electrons, H$^{+}$ (proton or hydrogen ion),  He$^{+}$ (singly ionized helium ion), and O$^{+}$ (singly ionized oxygen ion) ion species. The phenomenon of self-organization is widely observed in several systems, including magnetoplasmas. These magnetoplasmas exhibit a tendency to spontaneously organize themselves into ordered states. During the phenomenon of magnetic self-organization in plasmas, the plasma tends to minimize its energy while conforming to certain topological constraints. The process of magnetic self-organization in plasmas is also commonly referred to as relaxation \cite{Hasegawa1985,Ortolani}. The seminal research carried out by Woltjer and Taylor has proven that magnetized plasmas can relax into force-free states. The manifestation of these relaxed states occurs when the field aligns with its vorticity, i.e., $\mathbf{\nabla}\times\mathbf{B}=\lambda\mathbf{B}$, where $\mathbf{B}$ is the magnetic field and $\lambda$ is a constant and it is eigenvalue of the curl operator, also termed the scale parameter. It is of significance to acknowledge that when the vortex of a vector aligns with the vector itself, it is referred to as a Beltrami field. Consequently, the aforementioned Woltjer-Taylor relaxed state is also commonly known as a Beltrami state. The explanation for the observed magnetic field structures in the reversed field pinch phenomenon was provided through the Woltjer-Taylor concept of relaxation. However, this relaxation theory lacks the incorporation of crucial elements such as flow and pressure gradients, which are fundamental characteristics of practical plasmas \cite{Woltjer1958,Taylor1974,Taylor1986}. 

In order to derive the relaxed states that exhibit significant flow and pressure gradients, the theory of relaxation for a single fluid plasma was extended to include two fluid plasmas.  Research has demonstrated that a non-force-free double Beltrami state (DB) describes the self-organized state of two fluid plasmas composed of ions and inertialess electrons. This DB non-force-free state is a linear combination of two single Beltrami states, is defined by two self-organized vortices, and is the consequence of a strong coupling between magnetic fields and velocities. Utilizing the variational principle, the DB state can also be obtained by minimizing the total magnetofluid energy under the topological constraints of magnetic helicity for inertialess electron species and generalized helicity for ion species \cite{Steinhauer1997,Mahajan1998,Yoshida2002,Steinhauer2002}. Importantly, these Beltrami states in two fluid plasmas have been used extensively to model various plasma phenomena in astrophysical environments, such as heating of solar corona \cite{Mahajan2001}, flow generation in solar subcoronal regions \cite{Mahajan2002}, explosive or eruptive events in solar atmosphere \cite{Ohsaki2002,Kagan2010}, solar arcades \cite{Bhattacharyya2007}, turbulence in solar wind \cite{Krishan2005}, dynamo and reverse dynamo processes in astrophysical plasmas \cite{Mahajan2005,Lingam2015}. 

Further investigations have shown that the role of plasma species inertia is significant in the phenomenon of self-organization. It has been demonstrated that when the inertial effects of both species in a two-fluid plasma are considered, the index of Beltrami states increases. In the context of an electron-positron plasma, it has been observed that the relaxed state is a triple Beltrami (TB) state. The TB state is the linear superposition of three single Beltrami fields and is characterized by three relaxed state structures \cite{Bhattacharyya2003}. The diverse applications of Beltrami fields have inspired the scientific community to investigate and analyze the relaxed equilibrium states of plasmas containing more than two species. For instance, the relaxed state of three-component plasma, taking into account the inertia of all plasma species, is a quadruple Beltrami state (QB), which is a linear superposition of four Beltrami states and is characterized by four self-organized structures \cite{Shatashvili2016}.

In the context of the present study, it is imperative to emphasize that prior investigations on relaxed states in multi-ion plasmas were carried out employing two species of positive ions. In this regard, Shukla first obtained a DB state by assuming inertialess electrons and the same Beltrami parameters  (--a measure of generalized helicities of plasma species) for both ion species \cite{Shukla2005}. Following that, a recent study conducted by Gondal came to the conclusion that the relaxed state for inertialess electrons with different Beltrami parameters of inertial ion species is the TB state \cite{Gondal2022}. In scenarios where all plasma species are considered inertial, the relaxed state is the TB state, provided that the Beltrami parameters for the ion species are assumed to be same \cite{Iqbal2012}. In the most recent investigation by Ullah et al., it was found that the relaxed state is the QB state when all the plasma species are inertial and have different Beltrami parameters or generalized helicities \cite{Shafa2022}. Likewise, the study of Beltrami states has been carried out in electron-depleted dusty multi-ion plasmas, which consist of stationary negatively charged dust particles in addition to one positive ion species and two negative ion species. In this plasma model, multi-Beltrami relaxed states such as DB, TB, and QB have also been formulated and investigated by taking into consideration the inertial effects of both positive and negative ion species and adjusting the generalized vorticities of ion species \cite{Iqbal2014,Iqbal2015,Gondal2020}. Furthermore, it is also crucial to mention that in recent years, such multi-Beltrami states and their implications have also been explored in a variety of astrophysical plasmas, such as dense degenerate plasmas \cite{Shatashvili2016,Berezhiani2015,Shatashvili2019}, classical relativistic hot plasmas \cite{Usman2021,Usman2023a,Usman2023b,Usman2023c,Usman2024}, and general relativistic plasmas \cite{Asenjo2019,Bhattacharjee2020}. 

In this paper, a QB relaxed state for a four-component multi-ion plasma with inertialess electrons and inertial H$^{+}$, He$^{+}$, and O$^{+}$ ions is derived and investigated. The plasma parameters chosen for this investigation are based on in situ observations conducted in the inner magnetosphere \cite{Jahn2017}. The findings of the study indicate that generalized helicities as well as variations in the densities of ion species have the potential to transform the characteristics of QB structures. Additionally, the potential consequences of QB field structures for the inner magnetosphere of earth are pointed out. These potential implications include diamagnetic and paramagnetic trends, as well as heating effects caused by disparate length scales. Regarding the originality of this study, it is important to mention that, to the best of our knowledge, no prior research has been conducted on the relaxed states of the multi-ion plasma consisting of three positive ion species present in the inner magnetosphere of the earth.

The present study is structured in the following manner: In section \ref{S2}, from the model equations, a QB relaxed state is derived. In the next section, the characteristics of QB state scale parameters as well as an analytical solution to the QB equation in an axisymmetric cylindrical geometry is given, and the effect of ion species density on QB field structures is studied. In section \ref{S4}, a brief summary and conclusion of the present investigations are given.
\section{QB field equation}\label{S2}
Consider an incompressible magnetized quasineutral multispecies plasma consisting of inertialess electrons and inertial H$^{+}$, He$^{+}$, and O$^{+}$ ion species. The quasineutrality condition for the plasma model under consideration can be expressed as
\begin{equation}
n_{p}+n_{h}+n_{i}\approx n_{e},
\end{equation}
where the subscripts $\alpha=p$, $h$, $i$, and $e$ denote H$^{+}$, He$^{+}$, O$^{+}$, and electron species, respectively, while  $n_{\alpha}$ is the number density of $\alpha$ plasma species. The momentum balance equations for plasma species in dimensionless form can be expressed as
\begin{equation}
  \frac{\partial \mathbf{P}_{e}}{\partial t} =\mathbf{V}_{e}\times \mathbf{
\Omega }_{e}-\mathbf{\nabla }\Psi _{e},\label{me}  
\end{equation}
\begin{equation}
  \frac{\partial \mathbf{P}_{p}}{\partial t} =\mathbf{V}_{p}\times \mathbf{
\Omega }_{p}-\mathbf{\nabla }\Psi _{p},\label{mp}
\end{equation}
\begin{equation}
  \frac{\partial \mathbf{P}_{h}}{\partial t} =\mathbf{V}_{h}\times \mathbf{
\Omega }_{h}-\mathbf{\nabla }\Psi _{h},\label{mh}  
\end{equation}
\begin{equation}
  \frac{\partial \mathbf{P}_{i}}{\partial t} =\mathbf{V}_{i}\times \mathbf{
\Omega }_{i}-\mathbf{\nabla }\Psi _{i},\label{mi}  
\end{equation}
where $\mathbf{P}_{e}=\mathbf{A}$, $\mathbf{P}_{p}=\mathbf{V}_{p}+
\mathbf{A}$, $\mathbf{P}_{h }=\mathbf{V}_{h}+\left(
m_{p}/m_{h}\right)\mathbf{A}$, $\mathbf{P}_{i }=\mathbf{V}_{i}+\left(
m_{p}/m_{i}\right) \mathbf{A}$, $\mathbf{\Omega }_{e}=\mathbf{\nabla}
\times\mathbf{P}_{e}$, $\mathbf{\Omega }_{p}=\mathbf{\nabla}
\times\mathbf{P}_{p}$, $\mathbf{\Omega }_{h}=\mathbf{\nabla}
\times\mathbf{P}_{h}$, $\mathbf{\Omega }_{i}=\mathbf{\nabla}
\times\mathbf{P}_{i}$, $\Psi _{e}=\phi
-\chi^{-1}p_{e}$, $\Psi _{p}=\phi
+p_{p}+0.5V_{p}^{2}$, $\Psi _{h }=(m_{p}/m_{h})\phi
+(\rho_{p}/\rho_{h})p_{h}+0.5V_{h}^{2}$, $\Psi _{i }=(m_{p}/m_{i})\phi
+(\rho_{p}/\rho_{i})p_{i}+0.5V_{i}^{2}$,
$\chi =n_{e}/n_{p}$, $\rho_{p}=m_{p}n_{p}$, $\rho_{h}=m_{h}n_{h}$, and $\rho_{i}=m_{i}n_{i}$. 
In equations of motion (\ref{me}-\ref{mi}), $\mathbf{P}_{\alpha}$ ($\alpha=e$, $p$, $h$, and $i$), $\mathbf{\Omega }_{\alpha}$, $\mathbf{V }_{\alpha}$, and $\Psi _{\alpha}$ are the generalized momentum, generalized vorticity, velocity, and generalized potential of plasma species, respectively. While $\mathbf{A}$ and $\phi$ are vector magnetic and scalar electric potentials related with magnetic ($\mathbf{B}=\mathbf{\nabla}\times \mathbf{A}$) and electric ($\mathbf{E}=-\mathbf{\nabla}\phi-c^{-1}(\partial\mathbf{A}/\partial t)$, where $c$ is speed of light in vacuum)  fields. Moreover, $e$, $m_{\alpha}$, and $p_{\alpha}$ are amount of elementary charge, mass, and thermal pressure ($p_{\alpha}=n_{\alpha}T_{\alpha}$, where $n_{\alpha}$, $T_{\alpha}$ are plasma species density and temperature) of plasma species, respectively. It is essential to point out that in this study, in order to express all of the model equations in a dimensionless form, we have normalized length, magnetic field, velocity of plasma species, time, scalar electric potential, vector magnetic potential, and thermal pressure with proton skin depth ($\lambda_{p}=\sqrt{m_{p}c^{2}/4\pi n_{p}e^{2}}$), some arbitrary magnetic field $B_{0}$, Alfv\'{e}n speed ($v_{A}=B_{0}/\sqrt{4\pi m_{p}n_{p}}$), proton gyroperiod ($\lambda_{p}/v_{A}$), $B_{0}^{2}/4\pi n_{p}e$, $\lambda_{p}B_{0}$, and $B_{0}^{2}/4\pi$, respectively. Now, we use Ampere's law in order to couple the dynamics of plasma species, which, in normalized form, can be written as  
\begin{equation}
\mathbf{\nabla \times B}=\mathbf{V}_{p}+\varepsilon\mathbf{V}
_{h}+\xi\mathbf{V}_{i}-\chi\mathbf{V}_{e}.\label{al}
\end{equation}
where $\varepsilon =n_{h}/n_{p}$ and $\xi =n_{i}/n_{p}$. Notably, in Ampere's law (\ref{al}), the displacement current is neglected due to the non-relativistic flow of plasma species. Further, by taking the curl of the equations of motion (\ref{me}-\ref{mi}), the following vorticity evolution equations for plasma species are obtained
\begin{equation}
  \frac{\partial \mathbf{\Omega }_{e}}{\partial t} =\mathbf{\nabla \times }
\left( \mathbf{V}_{e}\times \mathbf{\Omega }_{e}\right),\label{ve}  
\end{equation}
\begin{equation}
  \frac{\partial \mathbf{\Omega }_{p}}{\partial t}=\mathbf{\nabla \times }
\left( \mathbf{V}_{i}\times \mathbf{\Omega }_{p}\right), 
\end{equation}
\begin{equation}
  \frac{\partial \mathbf{\Omega }_{h}}{\partial t}=\mathbf{\nabla \times }
\left( \mathbf{V}_{h}\times \mathbf{\Omega }_{h}\right),
\end{equation}
\begin{equation}
  \frac{\partial \mathbf{\Omega }_{i}}{\partial t}=\mathbf{\nabla \times }
\left( \mathbf{V}_{i}\times \mathbf{\Omega }_{i}\right).\label{vi}
\end{equation}
It can be observed from vortex dynamics equations (\ref{ve}-\ref{vi}) that gradient terms do not play a role in the evolution of generalized vorticities of plasma species.
Before moving on to the next step of derivation of the relaxed state equation, it is important to point out that in the framework of model equations (\ref{me}-\ref{vi}), the magnetic helicity ($h_{e}$) for inertialess electron species, the generalized helicities ($h_{p}$, $h_{h}$, and $h_{i}$) for inertial ion species, and the magnetofluid energy ($W_{mf}$) are the ideal invariants or constants of motion for this plasma model \cite{Mahajan2015}. These ideal invariants can be expressed by the following relations
\begin{equation}
   h_{e}=\frac{1}{2}\int_{v}\left(\mathbf{A}\cdot\mathbf{B}\right)dv,\label{he}
\end{equation}
\begin{equation}
   h_{p}=\frac{1}{2}\int_{v}\left(\mathbf{P}_{p}\cdot\mathbf{\Omega}_{p}\right)dv,
\end{equation}
\begin{equation}
   h_{h}=\frac{1}{2}\int_{v}\left(\mathbf{P}_{h}\cdot\mathbf{\Omega}_{h}\right)dv,
\end{equation}
\begin{equation}
   h_{i}=\frac{1}{2}\int_{v}\left(\mathbf{P}_{i}\cdot\mathbf{\Omega}_{i}\right)dv,\label{hi}
\end{equation}
\begin{equation}
  W_{mf}=\frac{1}{2}\int_{v}\left(B^{2}+V_{p}^{2}+\frac{\rho_{h}}{\rho_{p}}V_{h}^{2}+\frac{\rho_{i}}{\rho_{p}}V_{i}^{2}\right)dv,
\end{equation}
where $\int_{v}$  represents volume integral over bounded domain $\Pi$ ($\subset R^{3}$) and  $dv$ is volume element. 
At this juncture, it is imperative to acknowledge that our plasma system, as described by Eqs. (\ref{me}-\ref{mi}), permits the presence of general steady-state solutions with $ \mathbf{V}_{\alpha }\times\mathbf{\Omega }_{\alpha }=0$, subject to the constraint that $\mathbf{\nabla}\Psi_{\alpha}=0$, where $\Psi_{\alpha}$ represents the fluid energy densities of plasma species. So in order to attain an equilibrium state, the most elementary and fundamental solution to Eqs. (\ref{ve}-\ref{vi}) is to align the generalized vorticities ($\mathbf{\Omega }_{\alpha }\parallel \mathbf{V}_{\alpha}$) and their corresponding flows; this type of solution is known as the Beltrami condition. Therefore, the Beltrami conditions derived from Eqs. (\ref{ve}-\ref{vi}) can be expressed as
\begin{equation}
\mathbf{\Omega }_{e}=a_{e}\mathbf{V}_{e},\label{be}
\end{equation}
\begin{equation}
\mathbf{\Omega }_{p }=a_{p}\mathbf{V}_{p},\label{bp}
\end{equation}
\begin{equation}
\mathbf{\Omega }_{h}=a_{h}\mathbf{V
}_{h},\label{bh}
\end{equation}
\begin{equation}
\mathbf{\Omega }_{i}=a_{i}\mathbf{V
}_{i},\label{bi}
\end{equation}
where $a_{e}$, $a_{p}$, $a_{h}$, and $a_{i}$ are called Beltrami parameters. Importantly, these Beltrami conditions (\ref{be}-\ref{bi}) also encapsulate fundamental physics principles, specifically, the concept that electrons devoid of inertia adhere to the magnetic field lines, while ions due to their inertia conform to the field lines that have been modified by their flow vorticity. Also the Beltrami parameters, denoted as $a_{e}$, $a_{p}$, $a_{h}$, and $a_{i}$, are determined through the measurement of magnetic and generalized helicities for ion species given by Eqs. (\ref{he}-\ref{hi}). In the case of slowly evolving systems, these values remain constant and serve as ideal invariants of this plasma model \cite{Mahajan2001,Ohsaki2001}. It is also important to note that as the Beltrami conditions (\ref{be}-\ref{bi}) are steady-state solutions of equations of motion (\ref{me}-\ref{mi}), given that the gradient terms independently vanish. The condition $\mathbf{\nabla}\Psi_{\alpha}=0$ leads to $\Psi_{\alpha}=$ constant, is referred to as Bernoulli's condition. The utilization of a Bernoulli conditions also enables the transformation of thermal energy into kinetic energy, as well as the redistribution of kinetic energy from a plasma characterized by high density and low velocity to a plasma characterized by low density and high velocity. So, the quasi equilibrium state characterized by the Beltrami and Bernoulli conditions is commonly known as the Beltrami-Bernoulli equilibrium state \cite{Mahajan1998}.

At this point, in order to obtain a relaxed state equation for the magnetic field, it is required to solve a system of equations given by Beltrami conditions (\ref{be}-\ref{bi}) along with Ampere's law (\ref{al}) simultaneously. Moreover, these equations also enable us to get the expressions for the velocities of ion species and the bulk fluid velocity in terms of the magnetic field. By eliminating $\mathbf{V}_{e}$, $\mathbf{V}_{h}$, and $\mathbf{V}_{i}$ from Eqs. (\ref{al},\ref{be}-\ref{bi}), the resulting expression for $\mathbf{V}_{p}$ can be stated as
\begin{equation}
\mathbf{V}_{p}=p_{1}\mathbf{\nabla}\times\mathbf{\nabla}\times\mathbf{\nabla}\times\mathbf{B}
-p_{2}\mathbf{\nabla}\times\mathbf{\nabla}\times\mathbf{B}+p_{3}\mathbf{
\nabla \times B}-p_{4}\mathbf{B},\label{vp}
\end{equation}
where $p_{1}=[( a_{p}-a_{h}) ( a_{p}-a_{i})
] ^{-1}$, $p_{2}=p_{1}( a_{h}+a_{i}-\eta ) $, $
p_{3}=p_{1}( 1+\mu _{h}+\mu _{i}+a_{h}a_{i}-\eta a_{h}-\eta
a_{i}) $, $p_{4}=p_{1}(a_{h}+a_{i}-a_{p}+\mu
_{h}a_{i}+\mu _{i}a_{h}-\eta a_{h}a_{i}) $, $\mu_{h}=\varepsilon m_{p}/m_{h}$, $\mu_{i}=\xi m_{p}/m_{i}$, and $\eta=\chi /a_{e}$. Likewise, an expression for $\mathbf{V}_{h}$ can be expressed as 
\begin{equation}
\mathbf{V}_{h}=h_{1}\mathbf{\nabla}\times\mathbf{\nabla}\times\mathbf{\nabla}\times\mathbf{B}
-h_{2}\mathbf{\nabla}\times\mathbf{\nabla}\times\mathbf{B}+h_{3}\mathbf{
\nabla \times B}-h_{4}\mathbf{B},\label{vh}
\end{equation}
where $h_{1}=[ \varepsilon ( a_{h}-a_{p})(
a_{h}-a_{i})] ^{-1}$, $h_{2}=h_{1}( a_{p}+a_{i}-\eta
) $, $h_{3}=h_{1}( 1+\mu _{h}+\mu _{i}+a_{i}a_{p}-\eta a_{i}-\eta
a_{p}) $, and $h_{4}=h_{1}( a_{i}+\mu _{h}a_{p}+\mu _{h}a_{i}+\mu
_{i}a_{p}-\mu _{h}a_{h}-\eta a_{i}a_{p}) $. In a similar fashion, the relation for $\mathbf{V}_{h}$ in terms of magnetic field can be written as 
\begin{equation}
\mathbf{V}_{i}=i_{1}\mathbf{\nabla}\times\mathbf{\nabla}\times\mathbf{\nabla}\times\mathbf{B}
-i_{2}\mathbf{\nabla}\times\mathbf{\nabla}\times\mathbf{B}+i_{3}\mathbf{
\nabla \times B}-i_{4}\mathbf{B},\label{vi}
\end{equation}
where $i_{1}=[ \xi ( a_{h}-a_{i})( a_{p}-a_{i})
] ^{-1}$, $i_{2}=i_{1}( a_{p}+a_{h}-\eta) $, $i_{3}
=i_{1}( 1+\mu _{h}+\mu _{i}+a_{p}a_{h}-\eta a_{p}-\eta a_{h}) $,
and $i_{4}=i_{1}( a_{h}+\mu _{h}a_{p}+\mu _{i}a_{p}+\mu _{i}a_{h}-\mu
_{i}a_{i}-\eta a_{h}a_{p}) $.  The expression for the bulk or composite plasma velocity, given by $\mathbf{V} =\rho_{t}^{-1}(\rho_{p}\mathbf{V}
_{p}+\rho_{h}\mathbf{V}
_{h}+\rho_{i}\mathbf{V}
_{i})$, can be obtained by utilizing the values of $\mathbf{V}_{p}$, $\mathbf{V}_{h}$, and $\mathbf{V}_{i}$ as provided in Eqs. (\ref{vp}-\ref{vi}). Specifically, the expression for $\mathbf{V}$ can be written as
\begin{equation}
\mathbf{V}=\nu _{1}\mathbf{\nabla}\times\mathbf{\nabla}\times\mathbf{\nabla}\times\mathbf{B}-\nu_{2}\mathbf{\nabla}\times\mathbf{\nabla}\times\mathbf{B}+\nu_{3}\mathbf{\nabla \times B}
-\nu_{4}\mathbf{B},\label{bv}
\end{equation}
where $\rho_{t}=\rho_{p}+\rho_{h}+\rho_{i}$, $\nu_{1}=\rho_{t}^{-1}(p_{1}\rho_{p}+h_{1}\rho_{h}+i_{1}\rho_{i})$, $\nu_{2}=\rho_{t}^{-1}(p_{2}\rho_{p}+h_{2}\rho_{h}+i_{2}\rho_{i})$, $\nu_{3}=\rho_{t}^{-1}(p_{3}\rho_{p}+h_{3}\rho_{h}+i_{3}\rho_{i})$, and $\nu_{4}=\rho_{t}^{-1}(p_{4}\rho_{p}+h_{4}\rho_{h}+i_{4}\rho_{i})$. Following the derivation of the plasma species flow velocities in relation to the magnetic field, our current objective is to derive the relaxed state equation for the magnetic field. This can be achieved by substituting the value of $\mathbf{V}_{i}$ obtained from Eq. (\ref{vi}) into Eq. (\ref{bi}), resulting in the equation
\begin{equation}
\mathbf{\nabla}\times\mathbf{\nabla}\times\mathbf{\nabla}\times\mathbf{\nabla}\times\mathbf{B}-\kappa _{1}\mathbf{\nabla}\times\mathbf{\nabla}\times\mathbf{\nabla}\times\mathbf{B}+\kappa _{2}\mathbf{\nabla}\times\mathbf{\nabla}\times\mathbf{B}-\kappa _{3}\mathbf{\nabla \times B}
+\kappa _{4}\mathbf{B}=0,\label{qb}
\end{equation}
where $\kappa _{1}=a_{p}+a_{h}+a_{i}-\eta $, $\kappa _{2}=1+\mu _{h}+\mu
_{i}+a_{p}a_{h}+a_{p}a_{i}+a_{h}a_{i}-\eta( a_{p}+a_{h}+a_{i})$, $\kappa _{3}=a_{h}+a_{i}+\mu _{h}( a_{p}+a_{i})+\mu_{i}(
a_{p}+a_{h})-\eta( a_{p}a_{h}+a_{p}a_{i}+a_{h}a_{i})
+a_{p}a_{h}a_{i}$, and $\kappa _{4}=a_{h}a_{i}+\mu _{h}a_{p}a_{i}+\mu
_{i}a_{p}a_{h}-\eta a_{p}a_{h}a_{i}$.
The Eq. (\ref{qb}), referred to as the QB relaxed state, can be mathematically represented as the combination of four distinct Beltrami fields. It is also important to realize that the expressions for plasma species velocities and composite velocity also show strong magnetofluid coupling; moreover, it can easily be proved that all vector fields in this plasma model represent QB fields. Hence, consideration of inertialess electron and inertial positive ion plasma species, in addition to the four distinct Beltrami parameters ($a_{e}$, $a_{p}$, $a_{h}$, and $a_{i}$) for each plasma species, leads to the emergence of the QB relaxed state in a four-component multi-ion plasma.

Notably, this plasma model reduces to a three component multi-ion plasma (e-H$^{+}$-O$^{+}$) in the absence of He$^{+}$ ions. Consequently, when contemplating inertialess electrons in a multi-ion plasma consisting of two positive ion species, a TB relaxed state is possible for distinct generalized helicities of the ion species \cite{Gondal2022}. In contrast, it is also possible to model a DB state for ion species with the same generalized helicities \cite{Shukla2005}. On the other hand, when the inertia of electron species is also taken into consideration, a TB state can be obtained by assuming that ion species have same generalized helicities \cite{Iqbal2012}, whereas a QB state can be derived when all plasma species have distinct generalized helicities \cite{Shafa2022}.
\section{Characteristics and analytical solution of QB states}\label{S3}
The QB field Eq.  (\ref{qb}) can be mathematically represented as a linear combination of four single Beltrami fields ($\mathbf{B}_{\alpha }$). This relaxed state is characterized by four scale parameters ($\lambda_{\alpha}$), which are the eigenvalues of four distinct force-free fields that adhere to the subsequent conditions:
\begin{eqnarray*}
\mathbf{\nabla}\times \mathbf{B}_{\alpha}=\lambda _{\alpha}\mathbf{B}_{\alpha }\dots(\text{in}\ \Pi),\\
\widehat{\mathbf{n}}\cdot\mathbf{B}_{\alpha}=0\dots(\text{on}\ \partial\Pi),
\end{eqnarray*}
where $\partial\Pi$ represents smooth boundary of $\Pi$ ($\subset R^{3}$) on which $\widehat{\mathbf{n}}$ is a unit normal vector.  The aforementioned scale parameters $\lambda_{\alpha}$ also serve as measures of shear or twist ($\lambda_{\alpha}=\mathbf{\nabla}\times \mathbf{B}_{\alpha}\cdot\mathbf{B}_{\alpha }/B_{\alpha }^{2}$), while their reciprocals provide a measure of the dimensions of the relaxed state structures ($\because [\lambda_{\alpha}]=[L^{-1}]$). More importantly, the commutative property of the curl operator enables us to write the QB Eq. (\ref{qb}) in terms of scale parameters as follows \cite{Yoshida1990}
\begin{equation}
\left( \text{curl}-\lambda _{1}\right) \left( \text{curl}-\lambda
_{2}\right) \left( \text{curl}-\lambda _{3}\right) \left( \text{curl}
-\lambda _{4}\right) \mathbf{B}=0.  \label{SQB}
\end{equation}
The coefficients $\kappa_{\alpha}$ of the QB Eq. (\ref{qb}) can be related to the eigenvalues $\lambda_{\alpha}$ of Eq. (\ref{SQB}) in the manner given below:
\begin{eqnarray}
\kappa_{1} &=&\lambda _{1}+\lambda _{2}+\lambda _{3}+\lambda _{4},\label{r1} \\
\kappa_{2} &=&\lambda _{1}\lambda _{2}+\lambda _{2}\lambda _{3}+\lambda
_{1}\lambda _{3}+\lambda _{1}\lambda _{4}+\lambda _{2}\lambda _{4}+\lambda
_{3}\lambda _{4}, \\
\kappa_{3} &=&\lambda _{1}\lambda _{2}\lambda _{3}+\lambda _{1}\lambda
_{2}\lambda _{4}+\lambda _{2}\lambda _{3}\lambda _{4}+\lambda _{1}\lambda
_{3}\lambda _{4}, \\
\kappa_{4} &=&\lambda _{1}\lambda _{2}\lambda _{3}\lambda _{4}.\label{r4} 
\end{eqnarray}
The aforementioned relations (\ref{r1}-\ref{r4}) between $\lambda_{\alpha}$ and $\kappa_{\alpha}$ also satisfy Vieta's formula, implying that the values of scale parameters correspond to the roots of the subsequent quartic equation
\begin{equation}
\lambda ^{4}-\kappa_{1}\lambda ^{3}+\kappa_{2}\lambda ^{2}-\kappa_{3}\lambda +\kappa_{4}=0.\label{qe}
\end{equation}
As discussed above, the QB field can be represented as a linear combination of four single and distinct Beltrami fields; hence, its analytical solution can also be described as a linear sum of four single Beltrami fields, i.e.,
\begin{equation}
\mathbf{B}=\sum\limits_{\alpha =1}^{4}C_{\alpha}\mathbf{B}_{\alpha}.
\end{equation}
where $C_{\alpha}$'s are constants that can be real or complex valued, and their values can be calculated by using some suitable boundary conditions. In this study, an axisymmetric cylindrical geometry is considered for the analysis of QB field and flow profiles. So the analytical solution for the QB field and flow in an axisymmetric cylindrical geometry can be mathematically represented as
\begin{equation}
\mathbf{B}=\sum\limits_{\alpha =1}^{4}C_{\alpha}[J_{1}\left( \lambda _{\alpha }r\right) 
\widehat{\theta}+J_{0}\left( \lambda _{\alpha }r\right)\widehat{z}],\label{AS1}
\end{equation}
\begin{equation}
\mathbf{V}=\sum\limits_{\alpha =1}^{4}C_{\alpha}f_{\alpha}[J_{1}\left( \lambda _{\alpha }r\right) 
\widehat{\theta}+J_{0}\left( \lambda _{\alpha }r\right)\widehat{z}],\label{AS2}
\end{equation}
where $f_{\alpha}=\nu_{1}\lambda^{3}_{\alpha}-\nu_{2}\lambda^{2}_{\alpha}+\nu_{3}\lambda_{\alpha}-\nu_{4}$.  In order to determine the values of $C_{\alpha}$'s in Eqs. (\ref{AS1}-\ref{AS2}), we use the following boundary conditions: $\left\vert \mathbf{B}_{z}\right\vert _{r=0}=b$, $\left\vert \mathbf{B}_{\theta }\right\vert _{r=d}
=f$, $\left\vert \left( \mathbf{\nabla }\times 
\mathbf{B}\right) _{z}\right\vert _{r=0}=g$, and $\left\vert \left( \mathbf{
\nabla }\times \mathbf{B}\right) _{\theta }\right\vert _{r=d}=s$, where $b$, $d$, $f$, $g$, and $s$ are some arbitrary real valued constants. From above these boundary conditions, one can get the following values for $C_{\alpha}$:
\begin{eqnarray*}
C_{1} &=&\Lambda(\gamma _{2}\left( \gamma _{4}\delta _{3}\vartheta _{5}-\gamma
_{3}\delta _{4}\vartheta _{4}-\vartheta _{6}\varrho _{2}\right) +\gamma
_{3}\left( \vartheta _{5}\varrho _{3}-\gamma _{4}\delta _{2}\vartheta
_{6}\right) -\gamma _{4}\vartheta _{4}\varrho _{4}), \\
C_{2} &=&\Lambda(\gamma _{4}\left( \vartheta _{2}\varrho _{4}+\gamma _{3}\delta
_{1}\vartheta _{6}-\gamma _{1}\delta _{3}\vartheta _{3}\right) +\gamma
_{1}\left( \gamma _{3}\delta _{4}\vartheta _{2}+\vartheta _{6}\varrho
_{1}\right) -\gamma _{3}\vartheta _{3}\varrho _{3}), \\
C_{3} &=&\Lambda(\gamma _{2}\left( \vartheta _{3}\varrho _{2}-\gamma _{4}\delta
_{1}\vartheta _{5}-\gamma _{1}\vartheta _{1}\delta _{4}\right) +\gamma
_{4}\left( \gamma _{1}\vartheta _{3}\delta _{2}-\vartheta _{1}\varrho
_{4}\right) -\gamma _{1}\vartheta _{5}\varrho _{1}), \\
C_{4} &=&\Lambda(\gamma _{1}\left( \gamma _{2}\delta _{3}\vartheta _{1}-\gamma
_{3}\delta _{2}\vartheta _{2}+\vartheta _{4}\varrho _{1}\right) +\gamma
_{3}\left( \gamma _{2}\delta _{1}\vartheta _{4}+\vartheta _{1}\varrho
_{3}\right) -\gamma _{2}\vartheta _{2}\varrho _{2}),
\end{eqnarray*}
where $\vartheta _{1}=\lambda _{1}-\lambda _{2}$, $\vartheta _{2}=\lambda
_{1}-\lambda _{3}$, $\vartheta _{3}=\lambda _{1}-\lambda _{4}$, $\vartheta
_{4}=\lambda _{2}-\lambda _{3}$, $\vartheta _{5}=\lambda _{2}-\lambda _{4}$, 
$\vartheta _{6}=\lambda _{3}-\lambda _{4}$, $\gamma _{\alpha }=J_{1}\left(
d\lambda _{\alpha }\right) $, $\delta _{\alpha }=g-b\lambda _{\alpha }$,  
$\varrho _{\alpha }=s-f\lambda _{\alpha }$, and $\Lambda=[\vartheta _{2}\vartheta _{5}( \gamma _{1}\gamma _{3}+\gamma
_{2}\gamma _{4}) -\vartheta _{1}\vartheta _{6}( \gamma _{1}\gamma
_{2}+\gamma _{3}\gamma _{4}) -\vartheta _{3}\vartheta _{4}(
\gamma _{2}\gamma _{3}+\gamma _{1}\gamma _{4})]^{-1}$.

After obtaining the analytical solution for the QB state, our main objective is to investigate the impact of plasma parameters on the relaxed state structures. The Eqs. (\ref{r1}-\ref{r4}) also illustrate that the eigenvalues are linked with the plasma parameters, such as Beltrami parameters, densities of ion species, and masses. The eigenvalues corresponding to the QB state can be classified as either purely real or as a combination of two real eigenvalues plus a pair of complex conjugate eigenvalues. A straightforward method for analyzing the nature of scale parameters involves utilizing the discriminant ($\Delta$) of the quartic Eq. (\ref{qe}). In the condition where $\Delta<0$, it can be observed that two scale parameters possess real values, while the other two scale parameters are complex conjugate pairs. In contrast, in cases where $\Delta>0$, it can be confirmed that all of the scale parameters possess either real values or exist as complex conjugate pairs. In particular, it can be noted that the eigenvalues exhibit the characteristics of being real and distinct when $\Delta>0$, $\Theta=8\kappa_{2}-3\kappa_{1}^{2}<0$, and $\Gamma=16\kappa_{1}^{2}\kappa_{2}-3\kappa_{1}^{4}-16\kappa_{1}\kappa_{3}-16\kappa_{2}^{2}+64\kappa_{4}<0$. Conversely, when $\Delta$, $\Theta$, and $\Gamma$ are all positive, the eigenvalues manifest as two pairs of complex conjugates.
\begin{figure}
    \centering
    \includegraphics{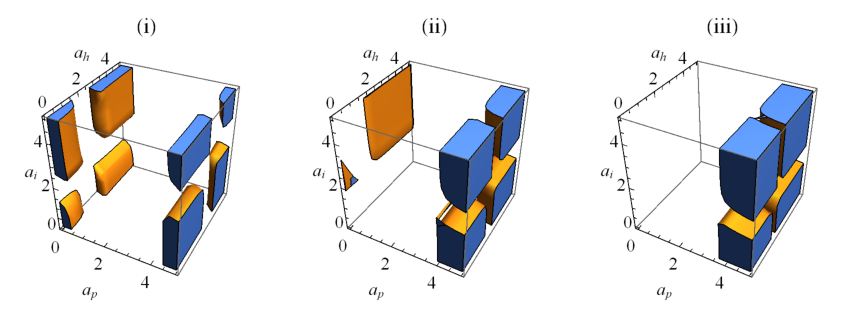}
    \caption{Character of QB state scale parameters as a function of Beltrami parameters $a_{p}$, $a_{h}$, and $a_{i}$ for $a_{e}=-1.0$, (i) $\chi=2$ , $\varepsilon=0.25$, and $\xi=1.25$, (ii)  $\chi=2$ , $\varepsilon=0.1$, and $\xi=0.9$, and (iii)  $\chi=2$ , $\varepsilon=0.03$, and $\xi=0.7$. In the colored region all the scale parameters are real and distinct.}
    \label{Fig1}
\end{figure}

Now, for the sake of investigation the QB relaxed state, we use the plasma parameters of the inner magnetosphere of the earth, where the ambient magnetic field and density of plasma species in $4-6$ $L-$shells are as follows: $B_{0}=1.44\times 10^{-3}-4.87\times 10^{-3}$ gauss, $n_{e}=10$ cm$^{-3}$, $n_{p}=4-5.8$ cm$^{-3}$, $n_{h}=0.2-1$ cm$^{-3}$, and $n_{i}=4-5$ cm$^{-3}$ \cite{Moya2022,Jahn2017}.
Firstly, in Fig. \ref{Fig1} (i-iii), the condition $\Delta>0$, $\Theta<0$, and $\Gamma<0$,  is plotted as a function of Beltrami parameters of ion species ($a_{p}$, $a_{h}$, and $a_{i}$) for the fixed value of  $a_{e}=-1.0$ and different ion species densities ((i) $\chi=2$ , $\varepsilon=0.25$, and $\xi=1.25$, (ii)  $\chi=2$ , $\varepsilon=0.1$, and $\xi=0.9$, and (iii)  $\chi=2$ , $\varepsilon=0.03$, and $\xi=0.7$) to show how Beltrami parameters and ion species density affect the nature of the scale parameters. In the colored regions of plots in Fig. \ref{Fig1}, all of the eigenvalues of the QB state are distinct and real, but in the transparent region, only two are real and the other two are complex conjugate pairs. It is evident from the plots in Fig. \ref{Fig1} (i-iii) that when the relative density of ion species varies, the nature of the scale parameters for given Beltrami parameter values also changes. Some of the real eigenvalues are changed into complex ones, and vice versa, as a consequence of a decrease in the densities of the He$^{+}$ and O$^{+}$ ion species. So, this change shows that the change in density of ion species in the inner magnetosphere of the earth will also cause a change in the trend of QB state equilibrium field and flow structures. For example, it can be observed from Eq. (\ref{AS1}) that in the case where all eigenvalues are real and under certain suitable boundary conditions, the relaxed state will exhibit paramagnetic features due to the decay of Bessel functions ($J_{0}$ and $J_{1}$) away from the center of the system.  In the case where two eigenvalues are real and the other two are complex conjugate pairs, the $J_{0}$ and $J_{1}$ functions transform into modified Bessel functions of the first kind ($I_{0}$ and $I_{1}$), which exhibit outward growth from the center. The combination of these ordinary and modified Bessel functions, when subjected to appropriate boundary conditions, leads to the emergence of diamagnetic or partially diamagnetic structures \cite{Mahajan1998}.

Since the relaxation events are commonly observed in plasmas within the earth's magnetosphere. These types of events refer to scenarios in which the magnetic topology of plasma undergoes a transformation, often accompanied by the release or transformation of energy. As a result, the plasma undergoes a transformation towards a state of minimum energy, leading to the possibility of a distinct topological configuration of the magnetic field compared to its initial condition. For instance, the substorm occurring in the magnetotail of the earth serves as a significant illustration of plasma relaxation; nevertheless, it should be noted that it is not the sole example. Empirical evidence suggests that, in the context of a substorm, there is a fast dissipation of energy within the magnetotail plasma. After this phase, the plasma in the magnetotail is relatively quiescent until the next period of relaxation \cite{Bhattacharjee1987,Paranicas1989}. Hence, the conversion between diamagnetic and paramagnetic structures in the relaxed state of the earth's inner magnetospheric multi-ion plasma offers the potential for energy conversion. This conversion involves the transformation of magnetic energy into kinetic energy and vice versa, which can be made possible by changes in generalized helicities and relative densities of ion species. Consequently, these energy transformation mechanisms associated with QB relaxed state structures can also play a significant role in the comprehension of substorm occurrences and plasma acceleration in the earth's magnetosphere.

\begin{figure}
    \centering
    \includegraphics{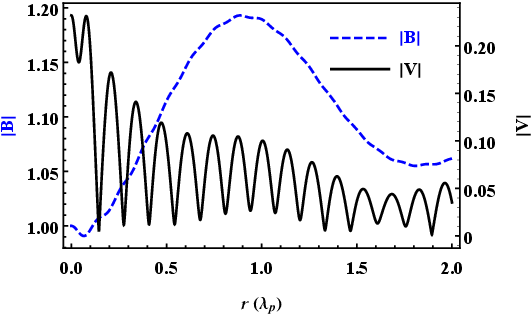}
    \caption{Magnetic field and flow profiles for $a_{e}=-40.0$, $a_{p}=4.0$, $a_{h}=4.5$, $a_{i}=51.2$, $\chi=2$ , $\varepsilon=0.1$, and $\xi=0.9$. }
    \label{Fig2}
\end{figure}
Afterwards, we demonstrate the ramifications of disparate variations in fields and flows resulting from disparate length scales of self-organized vortices. For this purpose, in the first case, we use the following values of plasma parameters and boundary conditions: $a_{e}=-40.0$, $a_{p}=4.0$, $a_{h}=4.5$, $a_{i}=51.2$, $\chi=2$ , $\varepsilon=0.1$, $\xi=0.9$, $b = 1.0$,  $f = 0.35$,  $g = 0.2$, and $s = 0.03$.
In accordance with the aforementioned plasma parameters, the eigenvalues of the QB state are $\lambda_{1}=0.3533$, $\lambda_{2}=3.7037$, $\lambda_{3}=4.3478$, and $\lambda_{4}=50.0$. Since the $\lambda_{\alpha}$ have the same dimensions as the reciprocal of length, the sizes of the relaxed state structures associated with these eigenvalues are $l_{1}=2.83\lambda_{p}$, $l_{2}=0.27\lambda_{p}$, $l_{3}=0.23\lambda_{p}$, and $l_{4}=0.02\lambda_{p}$, where $\lambda_{p}$ is the proton skin depth and it is equal to $1.14\times10^{7}$cm for $n_{p}=4$cm$^{-3}$. The aforementioned values indicate that only one self-organized structure possesses dimensions comparable to the proton skin depth, while the remaining three structures have dimensions that are smaller than the proton skin depth. So, in order to illustrate the field and flow patterns for these relaxed state structures, Fig. \ref{Fig2} is plotted. The plot demonstrates that a jittery and relatively weak flow is correlated with a relatively strong and smooth magnetic field. The presence of a flow field that exhibits fast and short scale variations suggests the presence of viscous damping associated with equilibrium structures. Hence, these self-organized field and flow structures have the potential to contribute to the heating of magnetospheric plasma. It is extremely important to point out that the inclusion of viscous terms in the equation of motion is omitted in our equilibrium analysis  \cite{Mahajan2001}.

\begin{figure}
    \centering
    \includegraphics{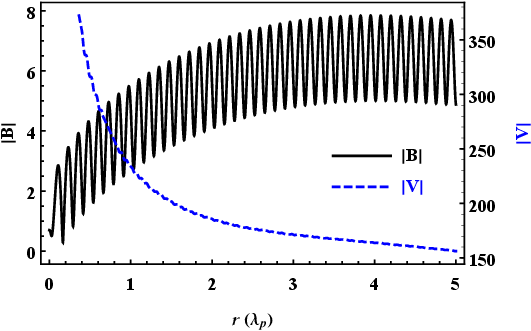}
    \caption{Magnetic field and flow profiles for $a_{e}=-50.0$, $a_{p}=50.1$, $a_{h}=50.5$, $a_{i}=51.0$, $\chi=2$ , $\varepsilon=0.1$, and $\xi=0.9$.}
    \label{Fig3}
\end{figure}
Contrary to the scenario discussed previously, when $a_{e}=-50.0$, $a_{p}=50.1$, $a_{h}=50.5$, $a_{i}=51.0$, $b = 0.7$,  $f = 0.25$,  $g = 0.2$, and $s = 0.03$, while the density of plasma species is kept the same, Fig. \ref{Fig3} demonstrates that a relatively weak and jittery magnetic field is associated with a robust and smooth flow. The eigenvalues along with the sizes of the self-organized structures are as follows: $\lambda_{1}=0.073$, $\lambda_{2}=50.1$, $\lambda_{3}=50.5$, $\lambda_{4}=51.0$, $l_{1}=13.74\lambda_{p}$, and $l_{2,3,4}\approx0.02\lambda_{p}$. These values indicate that one structure is significantly larger than $\lambda_{p}$, while the other three structures are considerably smaller than the $\lambda_{p}$. The observed phenomenon of a smooth and robust flow accompanied by a rather fast-varying weak magnetic field can be attributed to the presence of self-organized vortices of disparate sizes. This trend in the magnetic field and the flow profiles suggests the presence of resistive dissipation in the plasma. It is worth noting that our model equations do not explicitly account for such resistive and viscous dissipation effects.
\section{Conclusion}\label{S4}
The present study provides a theoretical framework for the self-organization of a multi-ion plasma consisting of electrons, H$^{+}$, He$^{+}$, and O$^{+}$ ion species.  In this plasma model, it is considered that all species are mobile, but the inertial effects of electrons are ignored. By using Ampere's law and the steady-state solution of vorticity evolution equations, which yield Beltrami conditions for plasma species, a QB relaxed state equation is derived. The QB state is non-force-free state, exhibits significant magnetofluid coupling, and is a linear superposition of four distinct single Beltrami states. In addition, it is characterized by four scale parameters, resulting in the formation of four self-organized structures. Taking into account the plasma parameters of the inner magnetosphere of the earth, the analysis demonstrates that the nature of the scale parameters can be transformed by varying ion species densities and plasma species' generalized helicities. These shifts in the nature of the scale parameters can convert paramagnetic structures into diamagnetic ones and vice versa. Such diamagnetic and paramagnetic trends in the QB state allow for plasma confinement as well as the possibility of energy conversion mechanisms comparable to substorm occurrences in the magnetosphere. The investigations also demonstrate that the fields and flows can vary at disparate length scales because of the disparate length scales of the self-organized vortices. In this regard, when one structure is of the order of proton skin depth while others are smaller than proton skin depth, a spatially fast-varying weak flow is coupled with a strong and smooth magnetic field, allowing for the prospect of viscous dissipation in plasma. In contrast, when three microscale structures coexist with a structure that is much larger than the proton skin depth, a smooth and strong flow is accompanied by a weak and jittery magnetic field, and this process suggests resistive dissipation in plasma. Importantly, we have not incorporated viscous and resistive effects into our plasma model, but these disparate self-organized structures allow for the possibility of such processes in the QB relaxed state. In conclusion, the findings from this study on multi-ion plasma will contribute to a deeper comprehension of various plasma phenomena observed in planetary magnetospheres and laboratory settings.
\section*{Conflict of interest statement}
None of the authors have a conflict of interest to disclose.
\section*{Data Availability Statement}
Data sharing is not applicable to this article as no datasets were generated or analyzed during the current study.

\end{document}